\documentclass[twocolumn,showpacs,preprintnumbers,amsmath,amssymb]{revtex4}

\usepackage{graphicx}
\usepackage{amssymb}

\newcommand{\bra}[2] {\mbox{}_{#2}\langle #1 |}
\newcommand{\ket}[2] {| #1 \rangle_{#2}}
\newcommand{\braket}[2] {\langle #1 | #2 \rangle}

\newcommand{\cat} [3] {\ket{#1}{#3}{#2}\ket{{-}#1}{#3}}
\newcommand{\sqo} [1] {\hat{S}_r\ket{1}{#1}}
\newcommand{\sqot} [2] {\hat{S}_{#2}\ket{1}{#1}}
\newcommand{\sqov}[2] {\hat{S}_{#2}\ket{0}{#1}}

\begin{document}

\title{Teleportation using squeezed single photons}

\author{Agata M.~Bra{\'n}czyk}
\affiliation{Center for Quantum Computer Technology, Department of Physics, The University of Queensland,
QLD 4072, Australia}
\author{T. C. Ralph}
\affiliation{Center for Quantum Computer Technology, Department of Physics, The University of Queensland,
QLD 4072, Australia}
\date{24 June 2008}

\begin{abstract}
We present an analysis of squeezed single photon states as a resource for teleportation of coherent state qubits and propose proof-of-principle experiments for the demonstration of coherent state teleportation and entanglement swapping. We include an analysis of the squeezed vacuum as a simpler approximation to small-amplitude cat states. We also investigate the effects of imperfect sources and inefficient detection on the proposed experiments.
\end{abstract}

\pacs{03.67.Lx, 42.50.-p, 42.50.Xa}

\maketitle
\section{Introduction}\label{sec:intro}
Coherent state quantum computing (CSQC) \cite{Ralph2003,Lund2008} is an optical approach to quantum computing which relies solely on linear optics, state preparation and
measurement, rather than in-line optical nonlinearities. Unlike in single photon linear optical quantum computing (LOQC) \cite{Knill2001,Kok2007} -- where qubits are encoded in the
polarisation, path, frequency etc. of single photons -- in CSQC, qubits are encoded in the phase and amplitude of coherent
states. The qubit basis is defined as
$\ket{\textbf{0}}{}=\ket{\alpha}{}$ and
$\ket{\textbf{1}}{}=\ket{{-}\alpha}{}$. The basis is only approximately orthogonal where
$\braket{\alpha}{{-}\alpha}=e^{-2|\alpha|^2}$, however, for $|\alpha|>2$, the
overlap is practically zero ($<\nolinebreak4\times\nolinebreak10^{-4}$).

In both CQCT and LOQC, teleportation is used to implement gates \cite{Gottesman1999}. A key practical difference between the two schemes is that a simple teleportation scheme, with a high probability of success, exists for CSQC \cite{Enk2001,Jeong2001}, whilst simple LOQC teleportation only works with a $50\%$ success rate \cite{Bouwmeester1997}. This leads to a significant saving in the overheads for computation. 

It was thought that $\alpha \geq 2$ was needed
in order to implement the required gates for quantum computing, due to the non-orthogonality of coherent state qubits \cite{Lund2005}. More recently, however, Lund \emph{et al.}
\cite{Lund2008} presented a universal set of gates for quantum
computing which work even for small $\alpha$. In this scheme, the
size of the coherent state has no effect on the fidelity of the gate
-- only on the probability of success of the gate. Indeed, below a
certain $\alpha$, these gates could not be used for scalable quantum
computing, as the probability of success would be too low, nevertheless the probability of 
success can still be significantly greater than the LOQC bound of 50\%. 
Such heralded gates open the door to a range of exciting
possibilities for proof-of-principle experimental implementation of
coherent state quantum computing.

The difficulty in performing these (and previous) gates does not
arise from our inability to create large amplitude coherent states
-- these are very well approximated by the output of a laser. The
difficulty arises from our inability to create \emph{superpositions}
of coherent states with large $\alpha$. Such coherent state superpositions, known in the
literature as \emph{cat states}, have not yet been experimentally
realised.

However, small amplitude cat states (so called ``kitten states") can be
approximated -- in some cases, very well -- using photon-subtracted squeezed vacuum states \cite{Lund2004}. Such states have
already been experimentally demonstrated \cite{Ourjoumtsev2006,Neergaard-Nielsen2006}. We note that a photon-subtracted squeezed vacuum state is mathematically equivalent to a squeezed single photon state. These terms will be used interchangeably depending on context throughout this paper. In this paper, we will also examine the squeezed vacuum state as a less complicated way of generating cat state approximations. We will not discuss approximations to kitten states via homodyne post-selection \cite{Ourjoumtsev2007}. 

This paper is organised as follows. In section \ref{sec:tele} we revisit the coherent state teleportation protocol introduced in \cite{Enk2001,Jeong2001}. In section \ref{sec:approx} we discuss approximations to cat states, in particular, the squeezed single photon state. We then analyse how well these squeezed single photon states perform as resource states for the teleportation of arbitrary coherent state qubits in section \ref{sec:res}. In section \ref{sec:exp}, we propose experimentally realisable demonstrations of coherent state teleportation using squeezed single photons. In section \ref{sec:error}, we analyse the effect of imperfect state preparation and inefficient detection on the teleportation scheme before we conclude and discuss our results in section \ref{sec:disc}.

\section{Teleportation}\label{sec:tele}

\begin{figure}[t!]
  \begin{center}
   \includegraphics[width=0.4\textwidth]{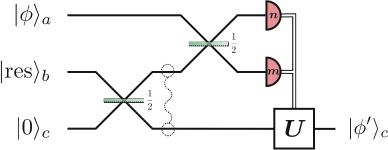}\\
  \end{center}
  \caption{Teleportation scheme for teleporting the state $\ket{\phi}{a}=\mu\ket{\alpha}{a}+\nu\ket{{-}\alpha}{a}$.
  A resource state, $\ket{\textrm{res}}{b}=\ket{\beta}{b}\pm\ket{{-}\beta}{b}$, incident on a $50:50$
  beamsplitter, creates a coherent-state Bell pair in modes $b$ and $c$. After the second beamsplitter, photon-number measurements of modes $a$ and $b$ project the state in mode $c$ into $\ket{\phi'}{c}=\ket{\phi}{a}$ (up to local unitaries). Depending on the measurement results, the remaining qubit may need
  to be corrected. The corrections (for an odd cat resource state) are as follows:  $n=\textrm{odd}$, $m=0$, $U\rightarrow I$; $n=0$, $m=\textrm{odd}$, $U\rightarrow
  X$;  $n=\textrm{even}$, $m=0$, $U\rightarrow Z$; $n=0$, $m=\textrm{even}$, $U\rightarrow
  XZ$, where $X$ and $Z$ are Pauli operators and $I$ is the
  identity operation. For experimental realisation of this scheme, $\ket{\textrm{res}}{b}=\sqot{b}{r}$ and $\ket{\phi}{a}=\ket{\alpha}{a}$, $\sqot{a}{r'}$ or $\hat{S}_{r''}\ket{0}{a}$, where $r=r_{\textrm{opt}}(\beta)$, $r'=r_{\textrm{opt}}(\alpha)$, $r''=r_{\textrm{opt-v}}(\alpha)$ and $\beta=\sqrt{2}\alpha$. The squiggly line emphasizes entanglement between qubits. }
  \label{fig:tele}
\end{figure}

The gates presented in \cite{Lund2008}  are all variations of the quantum teleportation scheme \cite{Enk2001,Jeong2001,Bennett1993} shown in figure \ref{fig:tele}. We will use this teleportation scheme as the basis for our proof-of-principle experiments. Refer to the caption in figure \ref{fig:tele} for a brief review of teleportation using cat-states.

A cat state is an equal coherent superposition of two coherent states $\ket{\beta}{}$ and  $\ket{{-}\beta}{}$, where:
 \begin{eqnarray}
\ket{\beta}{}=e^{-|\beta|^2/2}\sum^{\infty}_{n=0}\frac{\beta^n}{\sqrt{n!}}\ket{n}{}.
\end{eqnarray}
Using an odd cat $\cat{\beta}{-}{b}$ as the resource state, the combined input state can be written as follows (ignoring normalisations):

\begin{eqnarray}
\ket{\psi_{\textrm{in}}}{a,b,c}=\big(\mu\ket{\alpha}{a}{+}\nu\ket{{-}\alpha}{a}\big)\big(\cat{\beta}{-}{b}\big)\ket{0}{c}.
\end{eqnarray}
In this paper, we will use the convention that $\alpha$ refers to the initial amplitude of the input states in mode $a$ and $\beta$ refers to the initial amplitude of the resource states in mode $b$ and $\beta=\sqrt{2}\alpha$. Just before the photon
number measurement, the three-mode state is:
\begin{eqnarray}\label{eq:psi_out}\nonumber
\ket{\psi}{a,b,c}=\mu\ket{\beta,0,\alpha}{a,b,c}
&{-}&\mu\ket{0,\beta,{-}\alpha}{a,b,c}\\
+~\nu\ket{0,{-}\beta,\alpha}{a,b,c}
&{-}&\nu\ket{{-}\beta,0,{-}\alpha}{a,b,c}.
\end{eqnarray}

From equation (\ref{eq:psi_out}), it can be seen that photon
number measurements of modes $a$ and $b$ will project the state in mode $c$ into
$\mu\ket{\alpha}{c}+\nu\ket{{-}\alpha}{c}$, or some known variation
which can be corrected with single qubit Pauli operations (see figure \ref{fig:tele}). In practice, only the $X$
correction (which is simply implemented using a phase shifter) needs
to be performed for gate applications \cite{Jeong2007book}. Here we assume
corrections are done after detection of the output state. This is
called working in the ``Pauli frame'' \cite{Nielsen2005d}. 

\begin{figure}[t!]
  \begin{center}
   \includegraphics[width=0.45\textwidth]{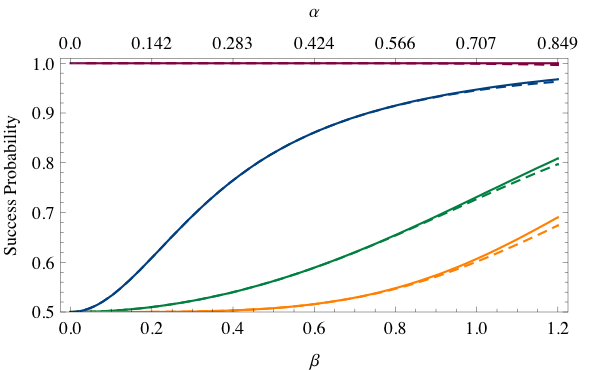}\\
  \end{center}
  \caption{Success probabilities of the teleportation scheme shown in figure
  \ref{fig:tele} using $\ket{\textrm{res}}{}=\cat{\beta}{-}{}$ (solid) and $\ket{\textrm{res}}{}=\sqo{}$ (dashed). The input states
  are (from top to bottom): $\frac{1}{\sqrt{2}}\ket{\alpha}{}-\frac{1}{\sqrt{2}}\ket{\alpha}{}$ (red);  $\frac{1}{2}\ket{\alpha}{}-\frac{\sqrt{3}}{2}\ket{\alpha}{}$ (blue); $\ket{\alpha}{}$ or $\ket{{-}\alpha}{}$ (green);
   and $\frac{1}{\sqrt{2}}\ket{\alpha}{}+\frac{1}{\sqrt{2}}\ket{\alpha}{}$ (yellow). Notice that the success probability is always larger than that of the LOQC scheme which can not succeed more than half of the time.}
  \label{fig:prob_succ_sq1_alpha}
\end{figure}

Ideally, it is not possible to detect a non-zero number of photons in both detectors simultaneously. The teleportation fails when both detectors get the result $0$. As
$\alpha$ increases, the probability of there being a $\ket{0}{}$ component in a coherent state decreases, therefore the probability of getting the result  $m=n=0$ also decreases, resulting in an increased probability of success. This
happens at different rates depending on the input state. In the
unique case of the input state being an odd cat state, there are
never any $m=n=0$ components and therefore the success
probability is always $1$. The solid curves in figure \ref{fig:prob_succ_sq1_alpha}
show the success probabilities for a selection of input states
which range between two extremities. When the gate succeeds, it does so with unity fidelity. 

\section{Approximating cat states}\label{sec:approx}

In this section, we will discuss experimentally realisable approximations to cat states. We will use the fidelity, $F(\rho,\ket{\phi}{})=\bra{\phi}{}\rho\ket{\phi}{}$, as a measure of how alike two states are, and therefore as a measure of how well these states approximate cat states. The fidelity ranges between $0$, where the states are orthogonal, and $1$, where the states are equal.

 Consider even and odd cat states of amplitude $\beta$ \cite{Walls1984}:
\begin{eqnarray}\nonumber
\ket{\textrm{even~cat}}{}&=&N_+(\ket{\beta}{}{+}\ket{{-}\beta}{})\\\label{eq:evencatstate}
&=&N_+e^{-\frac{1}{2}|\beta|^2}\sum_{n=0}^{\infty}\frac{2
\beta^{2n}}{\sqrt{2n!}}\ket{2n}{},\\\nonumber
\ket{\textrm{odd~cat}}{}&=&N_-(\ket{\beta}{}{-}\ket{{-}\beta}{})\\\label{eq:catstate}
&=&N_-e^{-\frac{1}{2}|\beta|^2}\sum_{n=0}^{\infty}\frac{2
\beta^{2n+1}}{\sqrt{(2n+1)!}}\ket{2n+1}{},
\end{eqnarray}
where $N_{\pm}=1/\sqrt{2(1\pm e^{-2|\beta|^2})}$. By writing the cat states in the Fock basis, we see that the even (odd) cat state contains only even (odd) photon number terms.

\begin{figure}[t!]
   \begin{center}
\hspace{1mm}\includegraphics[width=0.435\textwidth]{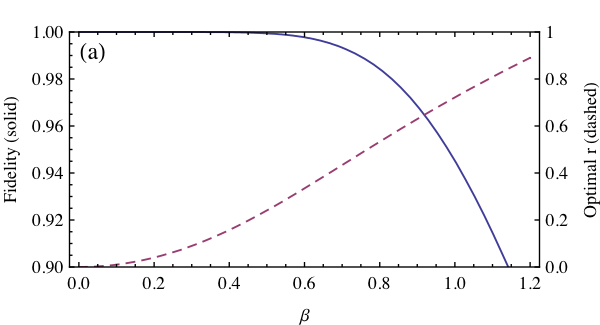}\\ \includegraphics[width=0.44\textwidth]{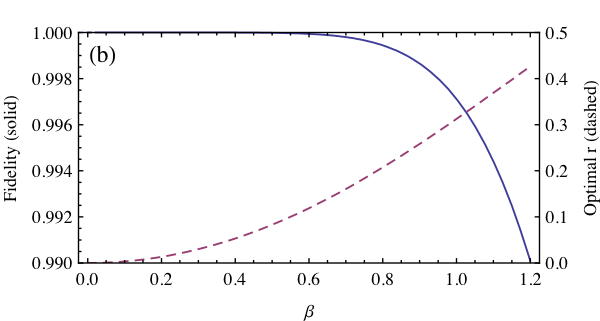}\\
  \end{center}
  \caption{(a) The fidelity between an even cat state and a squeezed vacuum state (solid)
  and the optimal squeezing parameter, $r$ (dashed), as a function of $\alpha$. (b) The fidelity between an odd cat state and a squeezed single photon (solid)
  and the optimal squeezing parameter, $r$ (dashed), as a function of $\alpha$. Notice different scales for the fidelity and the optimum squeezing parameter. }
  \label{fig:joint_fid_optr}
\end{figure}

The squeezed vacuum state:
\begin{eqnarray}\label{eq:sqone}
\hat{S}_{r}\ket{0}{}&=&\sum_{n=0}^{\infty}\frac{(\textrm{tanh}r)^n}{\sqrt{\textrm{cosh}r}}
\frac{\sqrt{2n!}}{2^nn!}\ket{2n}{},
\end{eqnarray}
where $r$ is the squeezing parameter, also contains only even photon number terms. The squeezed single photon is a Gaussian state, but nevertheless it is a high-fidelity ($F\nobreak>\nobreak0.99$) approximation to the small-amplitude ($\beta\nobreak\lessapprox\nobreak0.75$) even cat state. Optimising
over $r$, we find that the fidelity between $\ket{\beta}{}{+}\ket{{-}\beta}{}$ and $\hat{S}_{r}\ket{0}{}$ is at a maximum when $r_{\textrm{opt-v}}(\beta)=\textrm{Log}\Big(\sqrt{2\beta^2+\sqrt{1+4\beta^4}}\Big)$. Figure \ref{fig:joint_fid_optr} (a) shows how this fidelity and the
optimum amount of squeezing $r$ vary as a function of $\beta$. If one photon is subtracted from the squeezed vacuum state, the resulting state contains only odd photon number terms and is a high-fidelity ($F\nolinebreak>\nolinebreak0.99$) approximation to the small-amplitude ($\beta\nolinebreak\lessapprox\nolinebreak1.2$) odd cat state \cite{Dakna1997}. We remind the reader that this is also the squeezed single photon state. 
\begin{eqnarray}\label{eq:sqone}
\hat{a}\hat{S}_{r}\ket{0}{}=\hat{S}_{r}\ket{1}{}=\sum_{n=0}^{\infty}\frac{(\textrm{tanh}r)^n}{(\textrm{cosh}r)^{3/2}}
\frac{\sqrt{(2n+1)!}}{2^nn!}\ket{2n+1}{}.
\end{eqnarray}

\begin{figure}[b!]
  \begin{center}
\mbox{  \includegraphics[width=0.23\textwidth]{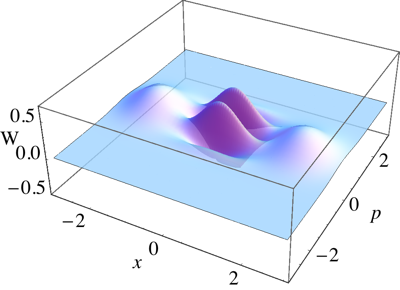}}
    \includegraphics[width=0.23\textwidth]{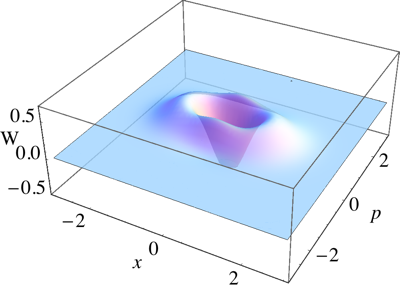}
      \includegraphics[width=0.23\textwidth]{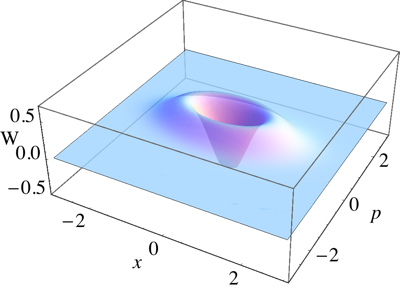}
      \caption{The Wigner function, $W$, where $x$ and $p$ are the in-phase and out-of-phase quadratures respectively, for: an odd cat state $\ket{\beta}{}{-}\ket{\beta}{}$ where $\beta=2$ (top-left); an odd cat state where $\beta=1$  (top-right); and a squeezed single photon $\hat{S}_r \ket{1}{}$ where $r=r_{\textrm{opt}}(1)\approx0.31$ (bottom). Notice that at $\beta=1$, the Wigner functions for the cat state looks very much like for the squeezed single photon. This becomes more pronounced at even lower $\beta$. In the limit of $\beta\rightarrow 0$, the odd cat state becomes an unsqueezed single photon. }
      \label{fig: threeplots}
  \end{center}
  \end{figure}

Figure \ref{fig: threeplots} shows the Wigner functions for odd cat states of amplitudes $\beta=2$ and $\beta=1$, as well as for a squeezed single photon.

Optimising
over $r$, we find that the fidelity between $\ket{\beta}{}{-}\ket{{-}\beta}{}$ and $\hat{S}_{r}\ket{1}{}$ is at a maximum when $r_{\textrm{opt}}(\beta)\nobreak=\nobreak\textrm{Log}(\sqrt{\frac{2\beta^2}{3}+\frac{1}{3}\sqrt{9+4\beta^4}})$.

Figure \ref{fig:joint_fid_optr} (b) shows how this fidelity and the
optimum amount of squeezing $r$ vary as a function of $\beta$. One can continue to subtract more photons, each time creating a better approximation to either an even or an odd cat state, however this very quickly becomes extraordinarily challenging to implement experimentally. A theoretical analysis of this method was performed in references \cite{Dakna1997,Nielsen2007}

In this paper, we focus on the squeezed single photon, as it is a better cat-state approximation than the squeezed vacuum state, and has already been experimentally demonstrated \cite{Ourjoumtsev2006,Neergaard-Nielsen2006}. However, we will include results for the squeezed vacuum state for comparison and to see just how well one can do with a gaussian state.  In the next section, we will investigate the eligibility of the squeezed single
photon state as an approximation to an odd cat state, for the
purposes of proof-of-principle implementation of the quantum teleportation described in \cite{Lund2008}. 

 We emphasize that it should not be taken for granted that a high-fidelity approximation to a cat state, will necessarily perform well in CSQC protocols. Take the example of the cat breeding protocol introduced by Lund \emph{et al.} and Jeong \emph{et al.} \cite{Lund2004, Jeong2005}. Using this scheme, it is possible to create larger cat states by interfering two smaller cat states on a beam splitter, then performing a measurement on one of the output modes. When a squeezed vacuum state is used as an approximation to an even cat state in this protocol, the resultant state is not the expected approximation to a larger cat state, but rather the same sized squeezed vacuum state that was input into the scheme. This is in spite of the high fidelity between the squeezed vacuum and the even cat state at small $\alpha$.

\section{Squeezed single photon as a resource}\label{sec:res}

\begin{figure}[b!]
  \begin{center}
        \includegraphics[width=0.44\textwidth]{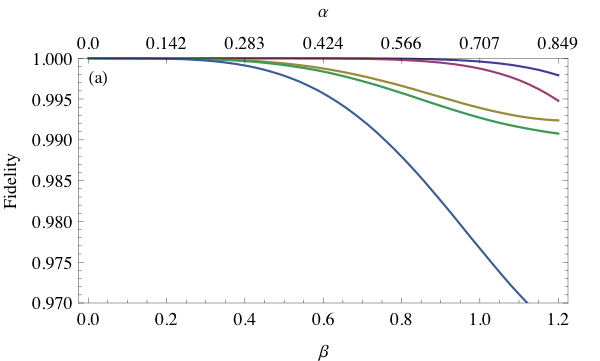}\\
   \includegraphics[width=0.44\textwidth]{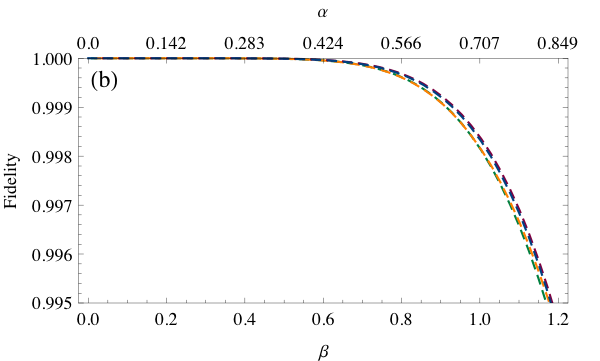}
  \end{center}
  \caption{(a) Fidelities for the teleportation of $\ket{\phi}{a}=\cat{\alpha}{-}{}$ using $\ket{\textrm{res}}{b}=\sqo{}$ given different photon-number measurement results, (from top to bottom) $m=1,2,3,4$ and $5$. (b) Fidelities for the teleportation scheme using $\ket{\textrm{res}}{b}=\sqo{}$ and a variety of input states (from top to bottom): $\ket{\alpha}{}-\ket{\alpha}{}$~(red);
   $\frac{1}{2}\ket{\alpha}{}+\frac{\sqrt{3}}{2}\ket{\alpha}{}$~(blue); $\ket{\alpha}{}+\ket{\alpha}{}$~(yellow); and $\ket{\alpha}{}$ or $\ket{{-}\alpha}{}$~(green). The input state has only a minor effect on the fidelity.}
  \label{fig:fid_dif_m}
\end{figure}

In this section, we will examine how well CS teleportation can be implemented using a squeezed single photon as a resource.  In section \ref{sec:approx}, we showed that the squeezed single photon is a very good approximation to a small-amplitude odd cat state (and that the squeezed vacuum is also a good approximation to an even cat state, but at smaller amplitudes).

To characterise how well the $\sqo{}$ theoretically performs as a resource for teleportation, we will calculate the fidelity  $F(\ket{\phi}{a},\ket{\phi'}{c})$ between the input state $\ket{\phi}{a}$ and the output state $\ket{\phi'}{c}$ . 

Since our resource state is only approximate, we will not have perfect interference at the beamsplitters. This will have three consequences. The first will be variations in the output state depending on the photon-number measurement results, as demonstrated in figure \ref{fig:fid_dif_m}(a). By taking into account the different possible output states and the probability with which we expect for them to occur, we can calculate the \emph{average} fidelity. One reason for this variation in fidelities, given different photon-number measurement results, is due to $r$ being optimised to maximise the fidelity between $\ket{\textrm{res}}{b}=\nobreak\sqo{}$ and $\ket{\textrm{res}}{b}=\cat{\beta}{-}{}$. This will maximise the average fidelity for the teleporter, but not the individual fidelities for each different photon-number result.  The second consequence of the imperfect interference will result in an input-state-dependent fidelity as demonstrated in \ref{fig:fid_dif_m}(b). This is only a small effect. Notice that the fidelity drops as a function of $\alpha$. This is solely due to the inadequacy of the $\sqo{}$ as an approximation to an odd cat state at high amplitudes, and not an artifact of the gate itself. 

The third consequence of the imperfect interference will be an additional way in which the gate can fail. Not only will it fail if we measure zero photons in both detectors, it will also fail if we measure a non-zero number of photons in both detectors \emph{simultaneously}, something which was not possible when we had perfect interference. This will result in a slightly decreased probability of success, which begins to manifest itself at larger amplitudes whereas the $m=n=0$ events are only problematic at low amplitudes. This is shown by the dashed curves in figure \ref{fig:prob_succ_sq1_alpha}.

One might expect that the the squeezed single photon will only be as good a resource as it is an approximation to an odd cat state. It is interesting to note that it is actually better.  A squeezed single photon $\sqo{}$ will have a certain fidelity when compared with an odd cat state $\ket{\beta}{}{-}\ket{{-}\beta}{}$, however, using that resource for teleporting $\ket{\phi}{a}=\ket{\alpha}{}{-}\ket{{-}\alpha}{}$ results in teleportation with a higher fidelity for a given $\beta$.

In the next section, we will look at teleporting physically realisable input states. 

\section{Proposed Experiments}\label{sec:exp}

In the previous section, we demonstrated that a squeezed single photon could be, in theory, a resource for teleportation of arbitrary superpositions of small-amplitude coherent states, however, at present, we are unable to create such superpositions. In this section we propose two types of experiment. The first is the teleportation of three particular examples states: a squeezed single photon as an approximation to an odd cat state; a squeezed vacuum as an approximation to an even cat state; and a coherent state. The second is an entanglement swapping scheme which demonstrates the effective teleportation of an arbitrary superposition of coherent states. In our calculations, the photon-number expansion of the states in this section were truncated at $n=15$, which was sufficient for accurate results up to $\beta=1.2$. 

\subsection{Teleportation}\label{sec:teleexp}

\begin{figure}[t!]
  \begin{center}
   \includegraphics[width=0.44\textwidth]{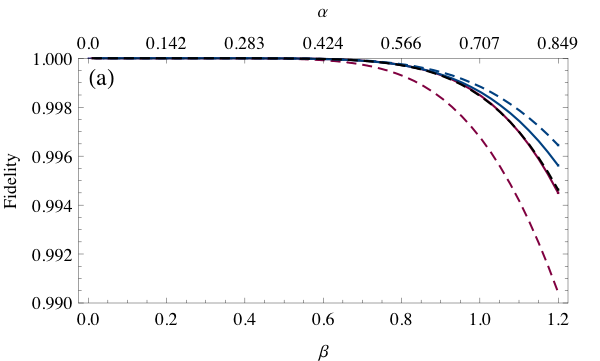}\\\hspace{0.5mm} \includegraphics[width=0.435\textwidth]{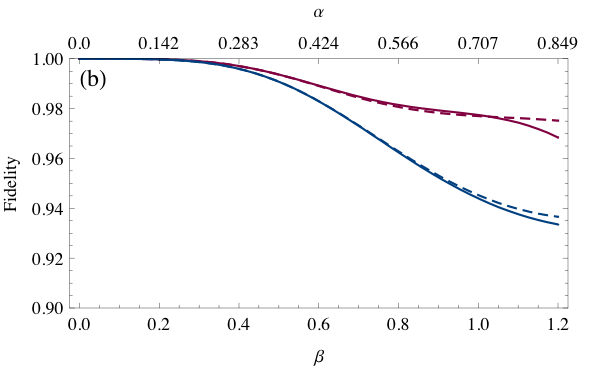}\\
  \end{center}
  \caption{(a) Fidelities for teleportation using $\ket{\textrm{res}}{b}=\sqot{r}{}$ and teleporting (solid, from top to bottom): $\ket{\phi}{a}=\cat{\alpha}{-}{}$; and $\ket{\phi}{a}=\cat{\alpha}{+}{}$ as well as: (dashed, from top to bottom)  $\ket{\phi}{a}=\sqot{a}{r'}$; $\ket{\phi}{a}=\ket{\alpha}{a}$; and $\ket{\phi}{a}=\sqov{a}{r''}$. (b) Fidelities for teleportation using $\ket{\textrm{res}}{b}=\sqov{a}{r}$ and teleporting (solid, from top to bottom): $\ket{\phi}{a}=\cat{\alpha}{+}{}$; and $\ket{\phi}{a}=\cat{\alpha}{-}{}$ as well as: (dashed, from top to bottom)  $\ket{\phi}{a}=\sqov{a}{r''}$; and $\ket{\phi}{a}=\sqot{a}{r'}$.  Notice different scales for the fidelity.}
  \label{fig:fidelity_compare}
\end{figure}

We will demonstrate CS teleportation by using the resource state $\ket{\textrm{res}}{}=\sqot{}{r}$ to teleport the following input states: a squeezed single photon $\sqot{}{r'}$ as an approximation to an odd cat state $\cat{\alpha}{-}{}$; a squeezed vacuum state $\sqov{}{r''}$ as an approximation to an even cat state $\cat{\alpha}{+}{}$; and a coherent state $\ket{\alpha}{}$.

To teleport a squeezed single photon $\sqot{}{r'}$ using another squeezed single photon $\sqot{}{r}$, we need to match the optimal squeezing parameters $r$ and $r'$. This can be achieved by relating the squeezed single photons to the odd cat states they are intended to approximate. This gives $r=r_{\textrm{opt}}(\beta)$ and $r'=r_{\textrm{opt}}(\alpha)$, where $\beta=\sqrt{2}\alpha$. We have calculated the fidelity averaged over only the odd photon-number results. For even results, a $Z$ correction is required, which would involve sending the output state through another gate making a meaningful comparison between the input and output states of the teleporter difficult.  Allowing for the $X$ correction is easy as it simply corresponds to a $\pi$ phase shift. All further fidelities shown in this section have also been averaged over odd photon-number results, for consistancy. We also calculated the fidelity for teleporting $\ket{\phi}{a}=\cat{\alpha}{-}{a}$ using $\ket{\textrm{res}}{b}=\sqot{b}{r}$. Both results are shown in figure \ref{fig:fidelity_compare} (a). For easy comparison with other figures, we have plotted the fidelity as a function of the effective $\beta$ for the squeezed single photon, rather than the squeezing parameter $r$. 

To teleport the squeezed vacuum state, we set $r''\nobreak=\nobreak r_{\textrm{opt-v}}(\alpha)$. The fidelities for teleporting $\sqov{}{r''}$ and $\cat{\alpha}{+}{}$ are also shown in figure \ref{fig:fidelity_compare} (a), as is the fidelity for teleporting $\ket{\phi}{}=\ket{\alpha}{}$. 

While there is some variation in the fidelity for the different input states, in the region where $\beta\lessapprox1.2$, the fidelity is always $>\nobreak0.99$, even when we teleport the squeezed vacuum state. 

Using the squeezed vacuum state as a resource for teleportation, however, does not do as well. This is because the resource state needs to be higher in amplitude than the input state and the squeezed vacuum is not as good an approximation to an even cat state at higher amplitudes. To achieve fidelities $>\nobreak0.99$, we can only use $\beta\lessapprox 0.5$. These results are shown in figure \ref{fig:fidelity_compare} (b). It is interesting to note that when using the squeezed vacuum state as a resource, the fidelities for teleportation do not follow the general trends of the fidelity between the ideal and approximate resource, as is the case with using the squeezed single photon as a resource. 

\begin{figure}[b!]
  \begin{center}
   \includegraphics[width=0.345\textwidth]{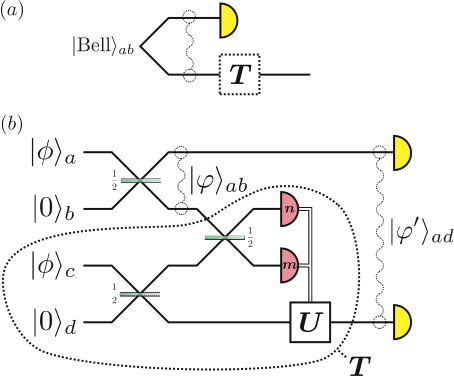}\\
  \end{center}
  \caption{(a) By measuring mode $a$ of a Bell state in an arbitrary basis, an arbitrary state can be prepared in mode $b$. This arbitrary state can be subsequently sent through a teleporter, $T$. (b) Replacing the teleporter $T$ with the teleporter described in figure \ref{fig:tele} and delaying the measurement of mode $a$, until \emph{after} the teleportation, results in an entanglement swapping scheme analogous to (a), where the
  Bell state is created by sending the state $\ket{\phi}{a}$ through a beam splitter and the teleporter
  consists of the circuit inside the dotted region.  By teleporting one qubit in the Bell state before measuring the other qubit in that Bell state, we are effectively teleporting all possible states. The homodyne measurements on modes $a$ and $d$ can be performed at the end in the form of state tomography. The squiggly lines represent entanglement between qubits.}
  \label{fig:ent swap}
\end{figure}

\subsection{Entanglement Swapping}\label{sec:entexp}
To truly demonstrate a teleportation protocol, one would like to show that the protocol is capable of teleporting an unknown arbitrary state. We will show how this can be done by casting the teleportation of an arbitrary unknown state into an entanglement swapping scheme. Refer to the caption in figure \ref{fig:ent swap} for details. 

\begin{figure}[b!]
  \begin{center}
   \includegraphics[width=0.44\textwidth]{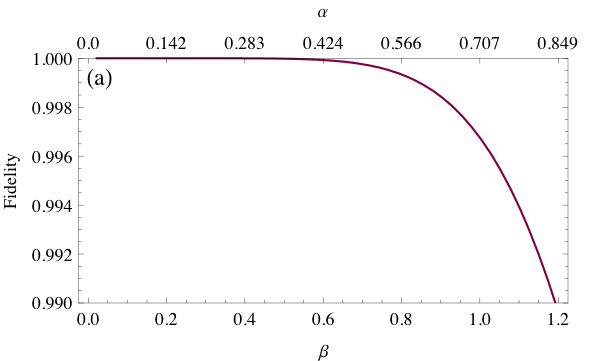}\\\hspace{0.5mm} \includegraphics[width=0.434\textwidth]{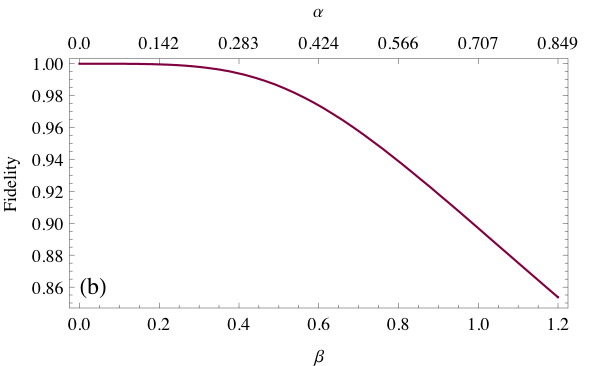}
  \end{center}
  \caption{Fidelity for an entanglement swapping protocol using: (a) $\ket{\phi}{}=\sqot{}{r'}$; and (b) $\ket{\phi}{}=\sqov{}{r''}$. Notice different scales for the fidelity. }
  \label{fig:prob_sq1_alpha_ent_swap}
\end{figure}

To characterise how well this protocol works, we will calculate the fidelity between the two-qubit entangled state in modes $a$ and $b$, after the first beamsplitter, and the two-qubit entangled state in modes $a$ and $d$, after the photon number measurements of modes $b$ and $c$. In our calculations, as with the teleportation scheme, we have omitted the even photon-number results. The average fidelity, over the odd photon-number results, is shown in figure \ref{fig:prob_sq1_alpha_ent_swap}. At first glance, it looks like the entanglement swapping protocol does not work as well as the teleportation scheme. This is because, for a given resource state of amplitude $\beta$, the input state to be teleported would have an amplitude $\alpha$ and the cat state approximations are much better at lower amplitudes. For the entanglement swapping scheme, we begin with two states of amplitude $\beta$, which means we are already starting with lower fidelity approximations. Nevertheless, for $\beta\lessapprox 1.2$, the fidelity is always $>0.99$. 

Using squeezed vacuum state to approximate an even cat state does not perform as well. To achieve fidelities of $>0.99$, we could only use $\beta\lessapprox0.45$.

\section{Error Analysis}\label{sec:error}

Until now, we have been treating the proposed experiments as lossless systems, however, propagation loss, imperfect detectors and loss in the source are likely to be issues in the experiment. In this section, we will investigate the effects of loss in the proposed experiment.

Errors could occur in a number of places: the state to be teleported and the resource state could be made imperfectly; there could be photon loss at the optical elements; and of course, inefficient photon number and homodyne detection. In this paper, we will assume that imperfect creation of the state to be teleported and any inefficiencies in the homodyne detection of the output state reflect our inability to analyse how well the scheme worked and are not fundamental to the scheme itself. These errors can be compensated for in the post measurement analysis of the data.  The calculations in this section were complicated by the additional loss modes, therefore, we needed to limit ourselves to smaller values of $\beta$. The photon-number expansion of the states in this section were truncated at $n=6$ for the teleportation scheme and $n=5$ for the entanglement swapping scheme which gave accurate results up to $\beta=1$ and $\beta=0.5$ respectively. 

\begin{figure}[b!]
  \begin{center}
   \includegraphics[width=0.40\textwidth]{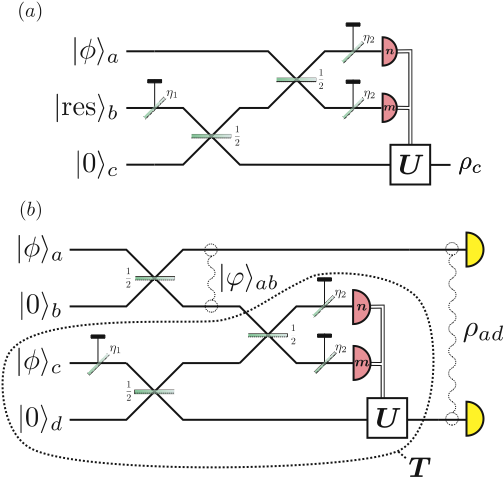} 
  \end{center}
  \caption{Teleportation scheme (a) and entanglement swapping scheme (b) with additional beamsplitters of transmitivity: $\eta_1$ to model the imperfect creation of the resourse state $\ket{\textrm{res}}{b}=\sqot{b}{r}$ and; $\eta_2$ to model the inefficiencies in the photo-detectors. The input state is $\ket{\phi}{a}=\sqot{a}{r'}$ where $r'=r_{\textrm{opt}}(\sqrt{\eta_1}\alpha)$ for the teleportation scheme and $r'=r_{\textrm{opt}}(\sqrt{\eta_1}\beta)$ for the entanglement swapping scheme  to match the amplitude of the lossy resource state.}
  \label{fig:error}
\end{figure}

\begin{figure*}[!t]
\begin{center} \includegraphics[height=0.205\textwidth]{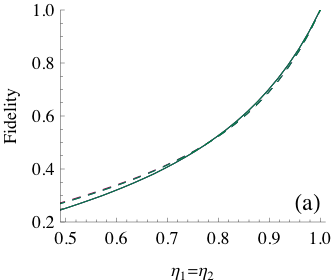}  \hspace{-2.3mm} \includegraphics[height=0.205\textwidth]{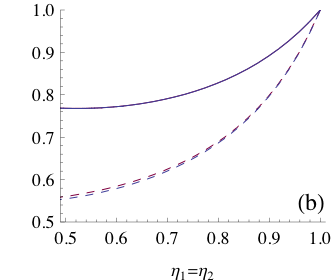} 
  \hspace{-2.3mm} \includegraphics[height=0.205\textwidth]{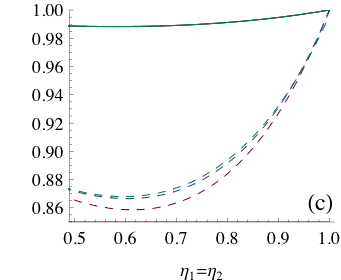} 
 \hspace{-2.3mm} \includegraphics[height=0.205\textwidth]{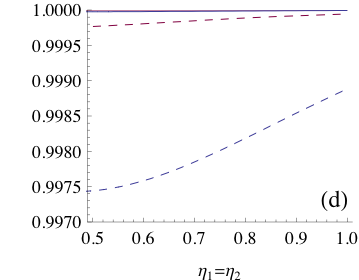}
 \end{center}
  \caption{The fidelity for the teleportation scheme as a function of $\eta_1$ and $\eta_2$ where $\eta_1=\eta_2$ for $\alpha=0.5$ (solid) and $\alpha=1.0$ (dashed). (a) $\ket{\textrm{res}}{}=\ket{\beta}{}{-}\ket{{-}\beta}{}$ and $\ket{\phi}{}=\ket{\alpha}{}{-}\ket{{-}\alpha}{}$ (red),  $\ket{\textrm{res}}{}=\sqot{}{r}$ and $\ket{\phi}{}=\ket{\alpha}{}{-}\ket{{-}\alpha}{}$ (blue) and $\ket{\textrm{res}}{}=\sqot{}{r}$ and $\ket{\phi}{}=\sqot{}{r'}$ (green). The fidelities for these states are so similar, they are practically indistinguishable. (b) From top to bottom, $\ket{\textrm{res}}{}=\ket{\beta}{}{-}\ket{{-}\beta}{}$ and $\ket{\phi}{}=\frac{1}{2}\ket{\alpha}{}{-}\frac{\sqrt{3}}{2}\ket{{-}\alpha}{}$ (red) and $\ket{\textrm{res}}{}=\sqot{}{r}$ and $\ket{\phi}{}=\frac{1}{2}\ket{\alpha}{}{-}\frac{\sqrt{3}}{2}\ket{{-}\alpha}{}$ (blue). For $\alpha=0.5$, the fidelities are also practically indistinguishable. (c) From top to bottom, $\ket{\textrm{res}}{}=\sqot{}{r}$ and $\ket{\phi}{}=\hat{S}_r\ket{0}{}$ (green),  $\ket{\textrm{res}}{}=\sqot{}{r}$ and $\ket{\phi}{}=\ket{\alpha}{}{+}\ket{{-}\alpha}{}$ (blue) and $\ket{\textrm{res}}{}=\ket{\beta}{}{-}\ket{{-}\beta}{}$  and $\ket{\phi}{}=\ket{\alpha}{}{+}\ket{{-}\alpha}{}$ (red). Again, practically indistinguishable at $\alpha=0.5$. (d) From top to bottom, $\ket{\textrm{res}}{}=\ket{\beta}{}{-}\ket{{-}\beta}{}$  and $\ket{\phi}{}=\ket{\alpha}{}$ (red) and  $\ket{\textrm{res}}{}=\sqot{}{r}$ and $\ket{\phi}{}=\ket{\alpha}{}$ (blue). Notice different scales for the fidelity.}
  \label{fig:Fid_loss}
\end{figure*}

\subsection{Teleportation}\label{sec:error_tele}

\begin{figure}[b!]
  \begin{center}
   \includegraphics[width=0.43\textwidth]{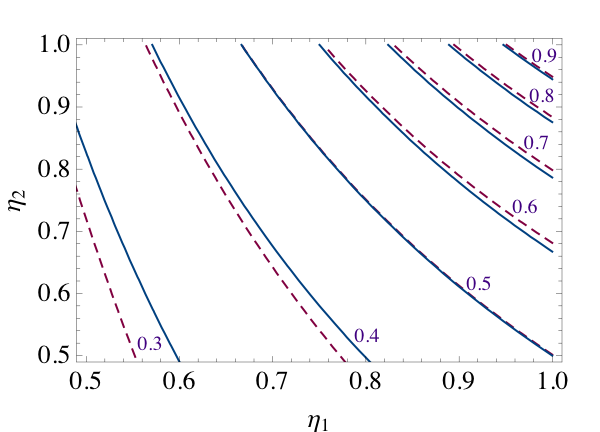} \\
       \includegraphics[width=0.43\textwidth]{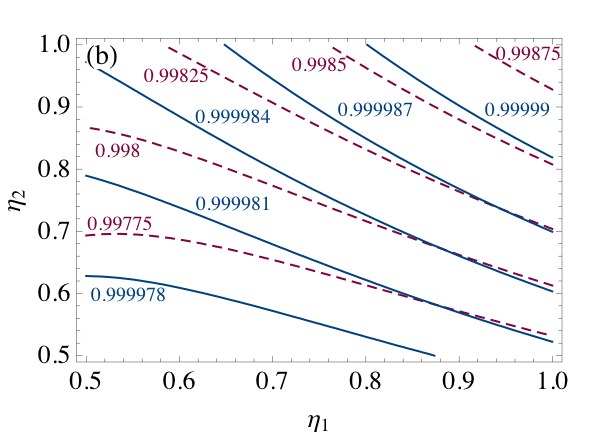} 
  \end{center}
  \caption{Contour plots of the fidelity for the teleportation scheme as a function of $\eta_1$ and $\eta_2$  for $\alpha=0.5$ (solid) and $\alpha=1.0$ (dashed): (a)  $\ket{\textrm{res}}{}=\sqot{}{r}$ and $\ket{\phi}{}=\sqot{}{r'}$; (b) $\ket{\textrm{res}}{}=\sqot{}{r}$ and $\ket{\phi}{}=\ket{\alpha}{}$.}
  \label{fig:error plot}
\end{figure}

We will model the imperfect creation of the resource state by placing a beamsplitter of transmitivity $\eta_1$ just after the source and the inefficient photo-detection by placing beamsplitters of transmitivity $\eta_2$ just before the detectors, as is shown in figure \ref{fig:error}. We assume that both detectors will have the same inefficiencies. $\eta$ ranges from $0$ to $1$ and at $\eta=1$, we have a lossless system. Our loss calculations have been carried out for the same scenario as in section \ref{sec:exp}, however, the loss in mode $b$ will decrease the amplitude of the resource state $\ket{\textrm{res}}{b}=\sqot{b}{r}$  by $\sqrt{\eta_1}$. To match this, our states to be teleported will need to be  $\ket{\phi}{a}=\ket{\sqrt{\eta_1}\alpha}{a}$, $\sqot{a}{r'}$ and $\sqov{a}{r''}$ where $r'=r_{\textrm{opt}}(\sqrt{\eta_1}\alpha)$ and $r''=r_{\textrm{opt-v}}(\sqrt{\eta_1}\alpha)$. Figure \ref{fig:error plot} shows the fidelity as a function of $\eta_1$ and $\eta_2$ using $\ket{\textrm{res}}{b}=\sqot{b}{r}$ and teleporting $\ket{\phi}{a}=\sqot{a}{r'}$ (a) and $\ket{\alpha}{a}$ (b) for $\beta=0.5$ (solid) and $\beta=1.0$ (dashed). As expected, loss has less effect on the fidelity at lower $\beta$, unfortunately, for $\ket{\phi}{a}=\sqot{a}{r'}$, decreasing $\beta$ does not improve the fidelity in the high-fidelity regime where one would like to perform experiments. Teleporting $\ket{\phi}{a}=\ket{\alpha}{a}$ is largely unaffected by loss. At both amplitudes, the fidelity remains extremely high. 

To compare these fidelities with those for other states, we have taken a slice through the contour plots at $\eta_1=\eta_2$ and plotted the fidelity as a function of the equal losses. This is shown in figure \ref{fig:Fid_loss}. We have calculated the fidelity for teleporting a selection of ideal input states using both an ideal odd cat state and the squeezed single photon approximation as resources. Where possible, we have also teleported approximations to the ideal input states. This was done for $\beta=0.5$ and $\beta=1.0$. This gives an idea of how much of the effect of loss is inherent to the scheme and how much is a result of the approximate input and resource states. At $\beta=0.5$, it is difficult to distinguish between the results for any of these variations between ideal and approximate input and resource states. At this amplitude, the effects of loss are fundamental to the scheme itself. It is interesting to note that for $\ket{\phi}{a}=\sqot{a}{r'}$ (or $\ket{\phi}{a}=\cat{\alpha}{-}{a}$), in the limit of $\eta_1\rightarrow 0$, the input state amplitude is matched such that $\alpha\rightarrow 0$, in which case, the input state becomes $\ket{\phi}{a}\rightarrow\ket{1}{a}$ and the output state will be $\ket{\phi'}{c}=\ket{0}{c}$ resulting in a fidelity which goes to $0$. However, for any other input state, as $\eta_1\rightarrow 0$, the input state will tend to $\ket{\phi}{a}\rightarrow\ket{0}{a}$. The output state will still be $\ket{\phi'}{c}=\ket{0}{c}$ resulting in a fidelity which goes to $1$. This is why the loss seems to have a greater effect on the fidelity for teleporting $\ket{\phi}{a}=\sqot{a}{r'}$ and $\ket{\phi}{a}=\cat{\alpha}{-}{a}$ than any other state. 

Increasing the amplitude to $\beta=1.0$ does not have much effect on teleporting the odd cat state (and its approximations), but it does decrease the fidelity for the other states. In attempting high-fidelity experiments, low loss levels will be essential when teleporting states similar to $\ket{\phi}{a}=\sqot{a}{r'}$ and $\ket{\phi}{a}=\cat{\alpha}{-}{a}$. States closer to $\ket{\phi}{a}=\cat{\alpha}{+}{a}$, $\ket{\phi}{a}=\ket{\alpha}{a}$ and their approximations are more forgiving.

\subsection{Entanglement Swapping}\label{sec:error_ent_swap}

\begin{figure}[h!]
  \begin{center}
   \includegraphics[width=0.415\textwidth]{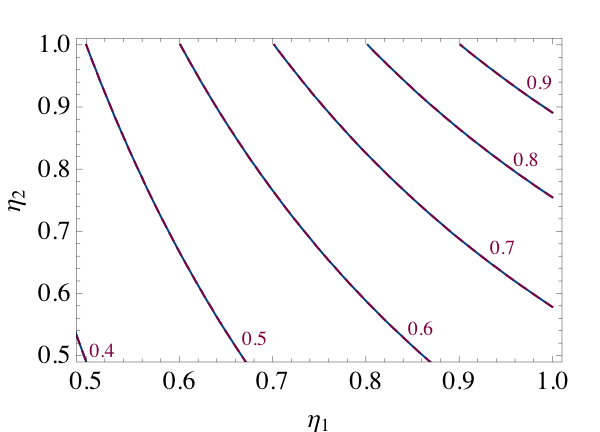} \\
       \includegraphics[width=0.415\textwidth]{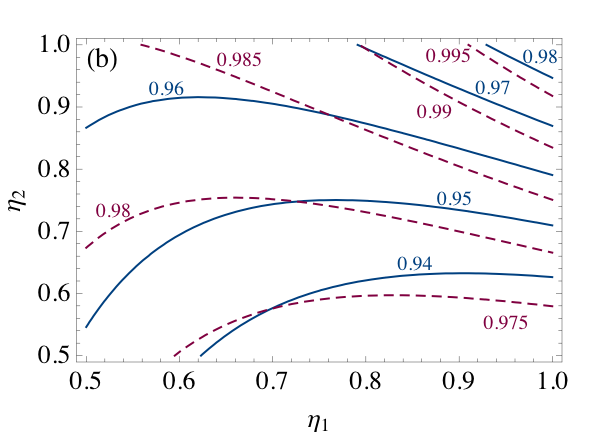} 
  \end{center}
  \caption{Contour plots of the fidelity for the entanglement swapping scheme as a function of $\eta_1$ and $\eta_2$ for $\beta=0.5$ where: (a) $\ket{\phi}{}=\cat{\alpha}{-}{}$ (dashed) and $\ket{\phi}{}=\sqot{}{r'}$ (solid); and (b) $\ket{\phi}{}=\cat{\alpha}{+}{}$ (dashed) and $\ket{\phi}{}=\sqov{}{r'}$ (solid).}
  \label{fig:contour_ent_swap}
\end{figure}

In the entanglement swapping protocol, we treat modes $a$ and $b$ as the input state to the teleporter and ignore loss in these modes in the analysis of the effects of loss on the teleporter, which has been modeled in much the same way as in section \ref{sec:error_tele}. We were only able to get reliable fidelities for $\beta=0.5$ as higher values of $\beta$ would require higher truncation of the photon number which, for a 7-mode calculation, was not computationally tractable. These fidelities are shown in figure \ref{fig:contour_ent_swap}.

At this amplitude, we can see that there is not much difference between performing the entanglement swapping with an ideal odd cat state or the squeezed single photon. Both are quite severely affected by loss. The ideal even cat is much more tolerant to loss whereas the squeezed vacuum is less so, however still better than the odd cat and squeezed single photon.

\section{Discussion}\label{sec:disc}

The squeezed single photon turns out to be a great resource for high-fidelity teleportation of small-amplitude coherent state superpositions. Due to its property of always containing at least one photon, the teleportation scheme will always succeed with a probability greater than $50\%$. We have shown that the squeezed single photon can be used to teleport a coherent state, a squeezed single photon and a squeezed vacuum, which despite not being a very good resource for teleportation itself, is a good approximation to an even cat state at very small amplitudes. In-principle teleportation of arbitrary coherent state superpositions can be demonstrated using the squeezed single photons as inputs to an entanglement swapping protocol. This also works with high fidelity at small amplitudes. Teleportation is the implementation of the identity gate and these results suggest that demonstration of more complicated non-trivial gates will be practical in the foreseeable future. 

Our analysis of the effects of imperfect source preparation and inefficient detection has shown this setup to be very fragile in this regard. It would be possible to do high-fidelity teleportation of states like the coherent state and the even cat state with a lossy system, but states which are more similar to the odd cat state degrade very quickly, even with low loss. It looks like this fragility is a property of the gate, and not just the approximation of the states, however, at higher  amplitudes, the fidelity is further affected by loss when using the approximate states. 

In this paper we have analysed coherent state teleportation using small-amplitude approximations to cat states, however, there has been progress in creating larger amplitude cat state approximations via cat state amplification \cite{Lund2004,Jeong2005} and ancilla-assisted photon-subtraction \cite{Takahashi2008}. 

We would like to thank A. Gilchrist for useful comments. This work is supported by the Australian Research Council.

\end{document}